\documentclass[prd,onecolumn,preprintnumbers,amsmath,amssymb,nofootinbib]{revtex4-2}

\usepackage{orcidlink}
\usepackage{graphicx}
\usepackage{dcolumn}
\usepackage{bm}
\usepackage{parskip}
\usepackage{subcaption}
\usepackage{slashed}
\usepackage{latexsym}
\usepackage[export]{adjustbox}
\usepackage{xcolor}

\usepackage[compat=1.1.0]{tikz-feynman}
\makeatletter
\pgfkeys{/graph drawing/.cd,
  node distance/.code       = \relax,
  level distance/.code      = \relax,
  sibling distance/.code    = \relax,
  horizontal/.code          = \relax,
  vertical/.code            = \relax,
  layered layout/.code      = \relax,
  spring layout/.code       = \relax,
}
\tikzfeynmanset{warn luatex=false}
\makeatother
\usepackage{hyperref}
\usepackage[capitalise]{cleveref}

\begin{document}

\title{Freeze-in and Freeze-out in a Right-Handed Neutrino Extended MSSM \\with a Seesaw Mechanism}
\author{Tushar Gupta\orcidlink{0009-0004-5961-6384}}
\email{tushar.gupta@helsinki.fi}
\affiliation{Department of Physics, and Helsinki Institute of Physics, University of Helsinki, P.O. Box 64, 00014 Helsinki, Finland}

\author{Matti Heikinheimo\orcidlink{0000-0003-0905-0488}}
\email{matti.heikinheimo@helsinki.fi}
\affiliation{Department of Physics, and Helsinki Institute of Physics, University of Helsinki, P.O. Box 64, 00014 Helsinki, Finland}

\author{Katri Huitu\orcidlink{0000-0002-7754-9320}}
\email{katri.huitu@helsinki.fi}
\affiliation{Department of Physics, and Helsinki Institute of Physics, University of Helsinki, P.O. Box 64, 00014 Helsinki, Finland}

\author{Harri Waltari\orcidlink{0000-0002-1626-1585}}
\email{harri.waltari@physics.uu.se}
\affiliation{Department of Physics and Astronomy, Uppsala University, P.O. Box 516, 75120 Uppsala, Sweden}

\date{\today}
\begin{abstract}
We investigate the possibility of saturating the relic density bound with light Higgsinos. When the minimal supersymmetric Standard Model is extended with right-handed neutrino superfields and the seesaw scale is very low, right-handed sneutrinos can be produced via the freeze-in mechanism. In such a case we can have essentially two independent sources for dark matter, the traditional freeze-out of Higgsinos and the freeze-in of right-handed sneutrinos. The heavier of these two will decay to the lighter species with a delay. We rule out such a scenario for all seesaw models as the lifetime of sterile neutrinos produced over-abundantly via Dodelson--Widrow mechanism exceeds the age of the universe and will contribute to the relic density.
\end{abstract}

\preprint{HIP-2025-30/TH}

\maketitle

\section{Introduction}

Supersymmetry is one of the most studied frameworks for beyond the Standard Model (SM) physics. Prior to the start of the Large Hadron Collider (LHC), it was often assumed that superpartners would be around the weak scale. However, the null results from numerous searches at the LHC have shown that this is not the case, at least for coloured superpartners. In fact, the discovery of a $125$~GeV Higgs boson points towards stop masses above a TeV \cite{Arbey:2011ab}.

Supersymmetry with R-parity conservation always provides a dark matter candidate, namely, the lightest supersymmetric particle (LSP). The most studied candidate is the lightest neutralino, which would act as a weakly interacting massive particle (WIMP). However, recent direct-detection experiments \cite{LZ:2024zvo,PandaX:2024qfu,XENON:2025vwd} have excluded nearly all of the parameter space for WIMP-like dark matter candidates. In addition, both the Higgsino and the wino lead to an underabundance unless their mass is about $1$~TeV for Higgsinos and $2$~TeV for winos \cite{Profumo:2004at}. This leads to some level of fine-tuning in the electroweak symmetry-breaking condition of the minimal supersymmetric Standard Model (MSSM):
\begin{equation}\label{eq:MSSMEWSB}
\frac{1}{2}m_{Z}^{2}=-\frac{m_{H_{u}}^{2}\tan^{2}\beta -m_{H_{d}}^{2}}{\tan^{2}\beta -1}-|\mu|^{2},
\end{equation}
where $m_{H_{u/d}}^{2}$ are soft supersymmetry-breaking masses and $|\mu|$ is essentially the Higgsino mass. \textit{A priori}, the supersymmetry-breaking parameters and the $\mu$ parameter are independent. They should therefore not be too far above the electroweak scale, otherwise large cancellations are needed between them. Hence, natural supersymmetry prefers a light Higgsino, while other superpartners can be heavier \cite{Baer:2013ava,Baer:2015rja}. This motivates the exploration of alternatives to WIMP dark matter.

An alternative way to produce dark matter is the freeze-in mechanism, where the dark matter candidate is so weakly coupled to other particles that it never attains thermal equilibrium \cite{Hall:2009bx,Bernal:2017kxu}. This requires the couplings between dark matter and the thermal bath to be of the order $10^{-7}$ or below. Such small couplings arise naturally in electroweak-scale seesaw models if the MSSM is extended with sterile-neutrino superfields. This could then create a possible link between neutrino physics and dark matter. For example, if the lightest sneutrino is lighter than the lightest superpartner in the thermal bath, there will always also be a freeze-out contribution, since the frozen-out particle would eventually decay to the lightest sterile-sneutrino state. A sterile sneutrino with tiny Yukawa couplings would evade all direct-detection constraints.

Such a scenario could be compatible with light Higgsinos, as suggested by the absence of fine-tuning in \cref{eq:MSSMEWSB}. In this framework, while the sterile sneutrino serves as the dark matter candidate, one could also consider the Higgsino playing that role with the freeze-in contribution from sterile sneutrinos giving an extra contribution to yield the correct amount of dark matter. However, this scenario is now practically ruled out by current direct-detection constraints \cite{LZ:2024zvo,PandaX:2024qfu,XENON:2025vwd} on the spin-independent DM nucleon scattering cross-section.

On the model side, there are four main options for adding the right-handed neutrino superfields. First, one can add them with the corresponding Yukawa couplings, leading to Dirac neutrinos. If we include the Majorana mass term for the right-handed neutrino, we get the type-I seesaw mechanism \cite{Minkowski:1977sc,Schechter:1980gr}. The other two possibilities involve two sets of sterile neutrinos, giving the linear \cite{Wyler:1982dd} and inverse seesaw \cite{Mohapatra:1986bd} models. In both cases, we have a small lepton number violating term, which makes the active neutrino masses small. By contrast, the seesaw mechanisms of type-II or type-III cannot produce freeze-in DM candidates as the (s)neutrinos have gauge interactions that lead to their thermalisation.

Feebly interacting right-handed sneutrinos have been studied as a dark matter candidate primarily in the case of Dirac neutrinos \cite{Asaka:2005cn,Asaka:2006fs,Biswas:2009zp,Banerjee:2018uut,Choi:2018vdi}. While most studies consider that the decays of the next-to-lightest supersymmetric particle (NLSP) provide the dominant source of dark matter, more recent work \cite{Choi:2018vdi} has pointed out that various $2\rightarrow 2$ scattering processes may contribute significantly, too.

In this work, we study models where right-handed neutrinos couple feebly to the MSSM and the lightest Higgsino mass lies at the electroweak scale, as suggested by naturalness arguments. We find that the case of Higgsino DM is severely constrained by direct-detection experiments. We show that the freeze-in scenario requires very light sterile neutrinos. These neutrinos are so long-lived that they themselves become a component of dark matter. However, in the region of the parameter space where the freeze-in mechanism does not lead to overabundance, sterile neutrino production via the Dodelson--Widrow mechanism \cite{Dodelson:1993je} will lead to overabundance in all simple SUSY seesaw models. Therefore, only the Dirac neutrino model, where sterile neutrinos mass eignestates are absent, can survive the relic density constraints \cite{Choi:2018vdi}. In \cite{Faber:2019mti} the authors studied a similar scenario than we do, but their scan resulted in a long-lived NLSP, which would affect Big Bang nucleosynthesis.

Our paper is organized as follows: In \cref{sec:model}, we review the MSSM extended with various seesaw mechanisms and constraints on them; \cref{sec:dmtheory} discusses the freeze-out and freeze-in processes generating dark matter and the properties of the dark matter candidates. In \cref{sec:results}, we outline our numerical procedure and present the results and then conclude in the following section.

\section{Seesaw extensions of the MSSM}\label{sec:model}

The freeze-in mechanism requires the DM candidate to remain out of thermal equilibrium. This implies that all its couplings to SM particles must be tiny, below $10^{-7}$ or so. Since, all gauge interactions would lead to rapid thermalisation, models in which (s)neutrinos are in a nontrivial representation of the SM gauge group cannot provide a freeze-in DM candidate. For this reason, we do not consider type-II and type-III seesaw mechanisms, where the MSSM field content is extended by SU(2) triplet superfields.

\subsection{Models with sterile (s)neutrinos}

The simplest of models with sterile-neutrino superfields is that of Dirac neutrinos. We extend the MSSM with three right-handed neutrino superfields $N^{c}_{i}$ (with lepton number $L=-1$), which are singlets under the SM gauge group. We write the superpotential in the form
\begin{equation}\label{eq:superpotential1}
W_{\mathrm{Dirac}}=y^{u}QH_{u}U^{c}+y^{d}QH_{d}D^{c}+y^{\ell}LH_{d}E^{c}
+\mu H_{u}H_{d}+y^{\nu}LH_{u}N^{c},
\end{equation}
where flavor indices are suppressed. The neutrino masses are then given by
\begin{equation}
m_{\nu,ij}=y^{\nu}_{ij}v\sin\beta,
\end{equation}
where $\tan\beta =v_{u}/v_{d}$ and $\langle H_{u,d}\rangle = v_{u,d}/\sqrt{2}$. While there is some uncertainty in the absolute scale of neutrino masses, we generally end up with $y^{\nu}\sim 10^{-13}$. This is the only model where neutrino masses are generated without lepton number violation.

Adding a Majorana mass term to this superpotential
\begin{equation}\label{eq:superpotential2}
W=W_{\mathrm{Dirac}}+\frac{1}{2}M_{Ni}N_{i}^{c}N_{i}^{c}
\end{equation}
results in the type-I seesaw model, where $M_{Ni}$ are significantly larger than the observed neutrino masses. The neutrino masses for a single generation are then given by the standard seesaw formula
\begin{equation}\label{eq:seesaw}
m_{\nu}\simeq \frac{(y^{\nu}v\sin\beta)^{2}}{2M},\quad m_{N}\simeq M.
\end{equation}

We consider a low seesaw scale, defined by
\begin{equation}
m_{\nu}\ll M_{Ni} \ll M_{W},
\end{equation}
which results in small neutrino Yukawa couplings $y^{\nu}\lesssim 10^{-10}$. The extreme case $M_{N}\rightarrow 0$ reduces to the Dirac neutrino scenario. Since the right-handed (s)neutrinos are coupled to other particles only via the neutrino Yukawa couplings, they decouple from the thermal bath in the early Universe, and they become rather long-lived or stable. Hence, their abundance is determined through the freeze-in mechanism.

Linear and inverse seesaw models are obtained by adding a second set of sterile-neutrino superfields $S_{i}$ (with lepton number $L=1$), resulting in the superpotentials
\begin{align}
W_{\mathrm{LS}} & = W_{\mathrm{Dirac}}+M_{N}N^{c}S+y^{LS}L\cdot H_{u}S,\label{eq:superpotential3}\\
W_{\mathrm{IS}} & = W_{\mathrm{Dirac}}+M_{N}N^{c}S+\frac{1}{2}\mu_{S}S^{2}.\label{eq:superpotential4}
\end{align}

In both models the last term is lepton number violating. The smallness of that term ensures the smallness of neutrino masses. In our case, we wish to have small Yukawa couplings for the (s)neutrinos, so we also assume that the seesaw scale, determined by $M_{N}$, is low. In principle, one could combine both of the superpotentials, but in practice one or the other seesaw mechanism dominates while the other one can be neglected.

The approximate formulae for light neutrino masses are
\begin{eqnarray}
m_{\nu} & \simeq & \frac{y^{\nu}y^{LS}v^{2}\sin^{2}\beta}{2M_{N}},\\
m_{\nu} & \simeq & \left( \frac{y^{\nu}v\sin\beta}{\sqrt{2}M_{N}}\right)^{2} \mu_{S}
\end{eqnarray}
for linear and inverse seesaw models, respectively. Compared to type-I seesaw model, these differ by factors of $y^{LS}/y^{\nu}$ and $\mu_{S}/M_{N}$ multiplying the mass formula.

As we shall discuss later, in the region of parameter space that corresponds to an acceptable relic density, the sterile neutrinos in the type-I seesaw model are so light that they become cosmologically stable and constitute an additional dark matter component. The relic abundance generated via the Dodelson--Widrow mechanism \cite{Dodelson:1993je} depends on the mixing angle between active and sterile neutrinos, and for the type-I seesaw model, we will show that sterile-neutrino production leads to overabundance. Therefore, we also study linear and inverse seesaw models, as they could suppress this mixing as the relation between active and sterile-neutrino masses is different. 

\subsection{Sneutrinos and their interactions}

In the presence of a seesaw mechanism, the mass matrix of the sneutrinos contains a lepton number violating part leading to a mass splitting between the CP-even and CP-odd sneutrino components \cite{Hirsch:1997is,Hirsch:1997vz}. In general, the sneutrinos are produced in states of a definite lepton number. If the mass splitting between the sneutrino components is small, both CP-components propagate coherently, and the state oscillates between a sneutrino state and an antisneutrino state \cite{Hirsch:1997is,Hirsch:1997vz,Grossman:1997is}, with the oscillation period being proportional to $\delta m^{2}$, the mass splitting between the CP-even and CP-odd sneutrino states.

In the type-I seesaw model, the sneutrino mass matrix for a single generation is of the form
\begin{equation}
M_{\nu,N}^{2}=
\begin{pmatrix}
0 & m_{\tilde{\ell}}^{2}+\frac{1}{8}(g^{2}+g^{\prime 2})v^{2}\cos 2\beta & 0 & A_{\nu}v_{u}-y^{\nu}\mu v_{d}\\
m_{\tilde{\ell}}^{2}+\frac{1}{8}(g^{2}+g^{\prime 2})v^{2}\cos 2\beta & 0 & A_{\nu}v_{u}-y^{\nu}\mu v_{d} & 0\\
0 & A_{\nu}v_{u}-y^{\nu}\mu v_{d} & M^{2} & m_{\tilde{N}}^{2}\\
A_{\nu}v_{u}-y^{\nu}\mu v_{d} & 0 &  m_{\tilde{N}}^{2} & M^{2}
\end{pmatrix}
\end{equation}
with the basis $(\tilde{\nu},\tilde{\nu}^{*},\tilde{N},\tilde{N}^{*})$ and neglecting terms $\mathcal{O}((y^{\nu})^{2})$. For Dirac neutrinos, $M^{2}=0$. Here $m_{\tilde{\ell}}^{2}$ and $m_{\tilde{N}}^{2}$ are the soft supersymmetry-breaking mass terms, and $A_{\nu}$ is the soft trilinear term. We assume that the soft masses $m_{\tilde{\ell}}^{2}$ and $m_{\tilde{N}}^{2}$ are much greater than all other terms, which leads to an extremely small mass splitting between the CP-even and CP-odd states generated by the lepton number violating mass term $M^{2}$. Therefore, a particle created as a (anti)sneutrino will remain nearly as an (anti)sneutrino for a long period. As the lepton number violating contribution is small in all our models (or zero in the case of Dirac neutrinos), the same argument applies for inverse and linear seesaw models. Hence, in the numerical procedure, we approximate the sneutrinos as eigenstates of definite lepton number; i.e., we define them as complex scalar fields.

The terms of the form $ A_{\nu}v_{u}-y^{\nu}\mu v_{d}$, which lie outside the $2\times 2$ block-diagonal parts of the mass matrix, mix left- and right-handed sneutrinos. In principle, this mixing could be large due to large A-terms \cite{Arkani-Hamed:2000oup}, but in that case right-handed sneutrinos would thermalise through the gauge interactions of the left-handed component. For simplicity, we assume that $A_{\nu}$ is vanishingly small, which also makes the mixing between left- and right-handed neutrinos negligible.

From the superpotential of \cref{eq:superpotential2,eq:superpotential3,eq:superpotential4}, we see that the right-handed sneutrinos only interact with Higgs bosons, Higgsinos, leptons and sleptons. The vertices between $\tilde{N}$ and $\tilde{H}^{\pm}\ell^{\mp}$, $\tilde{H}^{0}\nu$ are given by $y^{\nu}$, while in all but the linear seesaw model the vertices between the Higgs bosons and sneutrinos are
\begin{align}
\lambda_{H_{u}\tilde{N}\tilde{N}} & = -2i|y^{\nu}|^{2}v_{u},\label{eq:higgssnu1}\\
\lambda_{H_{u}\tilde{\nu}\tilde{N}} & = -i\sqrt{2}A_{\nu},\label{eq:higgssnu2}\\
\lambda_{H_{u}\tilde{\nu}\tilde{N}^{*}} & = -i\sqrt{2}y^{\nu}M,\label{eq:higgssnu3}\\
\lambda_{H_{d}\tilde{\nu}\tilde{N}} & = i\sqrt{2}y^{\nu}\mu,\label{eq:higgssnu4}
\end{align}
and the ones for charged Higgs bosons, charged sleptons, and right-handed sneutrinos are those of \cref{eq:higgssnu2,eq:higgssnu4} with an opposite sign. Since $y^{\nu}$ is extremely small and we assume $M\ll \mu$, we may neglect the interactions of \cref{eq:higgssnu1,eq:higgssnu3}.

In the linear seesaw model, \cref{eq:higgssnu1,eq:higgssnu3} get replaced by
\begin{eqnarray}
\lambda_{H_{u}\tilde{N}\tilde{N}} & = & -2i(|y^{\nu}|^{2}+|y^{LS}|^{2})v_{u},\\
\lambda_{H_{u}\tilde{\nu}\tilde{N}^{*}} & = & -i\sqrt{2}M(y^{\nu}+y^{LS}).
\end{eqnarray}
As alluded to above, the active-sterile mixing in this scenario becomes problematic because of the long lifetime of the sterile neutrino. In principle, one could increase $y^{LS}$ to achieve a viable neutrino spectrum with a heavier sterile neutrino in order to suppress the active-sterile mixing to an acceptable level for sterile-neutrino production. However, this would imply a larger coupling between the Higgs boson and sneutrinos, which increase the freeze-in contribution of sneutrinos and leads to overabundance. Hence, in the linear seesaw model, it is impossible to suppress the active-sterile mixing in the neutrino spectrum, keep a viable neutrino mass spectrum and satisfy the relic density constraints simultaneously. 

\section{Relic density and direct-detection properties}\label{sec:dmtheory}

\subsection{Freeze-out component}

In supersymmetric models with R-parity conservation, there is always some contribution of dark matter that is produced in thermal equilibrium through the freeze-out mechanism. In our case, motivated by \cref{eq:MSSMEWSB}, we take this to be a Higgsino-like neutralino, which is either stable or decays into right-handed sneutrinos after some delay.

In either case, the heavier of the two will eventually decay into the lighter species. However, due to the feeble interactions between the two sectors, the decay occurs with a delay. Typically, this delay is so long that by the time it happens, the number density of the lighter species has already diluted so much that the decay products no longer result in any further annihilations. As a result, the freeze-out and freeze-in processes remain independent.
For our parameter choices (see \cref{tab:parameters} below) the lifetime of the NLSP becomes of the order of $0.12$ seconds, so the energetic particles released in the decay of the NLSP have time to thermalise before Big Bang nucleosynthesis.

The neutralino mass matrix is
\begin{equation}\label{eq:Higgsinomass}
M_{\tilde{\chi}^{0}}=
\begin{pmatrix}
0 & \mu & -gv_{u}/\sqrt{2} & g^{\prime}v_{u}/\sqrt{2}\\
\mu & 0 & gv_{d}/\sqrt{2} & -g^{\prime}v_{d}/\sqrt{2}\\
-gv_{u}/\sqrt{2} & gv_{d}/\sqrt{2} & M_{2} & 0\\
g^{\prime}v_{u}/\sqrt{2} & -gv_{d}/\sqrt{2} & 0 & M_{1}\\
\end{pmatrix}
\end{equation}
in the basis $(\tilde{H}_{u}^{0},\tilde{H}_{d}^{0},\tilde{W}^{0},\tilde{B})$ with $v_{u}=v\sin\beta$ and $v_{d}=v\cos\beta$.

\begin{figure}
    \begin{center}
        \begin{tikzpicture}
            \begin{feynman}
                \vertex (a);
                \vertex [right = 2cm of a] (b);
                \vertex [left = 1.5cm of a] (c);
                \vertex [right = 1.5cm of b] (d);

                \vertex [above = 1.5cm of c] (i1) {\(\widetilde{\chi}^0\)};
                \vertex [below = 1.5cm of c] (i2) {\(\widetilde{\chi}^\pm\)};
                \vertex [above = 1.5cm of d] (f1) {\(SM\)};
                \vertex [below = 1.5cm of d] (f2) {\(SM\)};
            
                \diagram*{
                  (i1) -- [fermion] (a) -- [fermion] (i2),
                  (a)  -- [boson, edge label=\(W^\pm\)] (b),
                  (f1) -- [fermion] (b) -- [fermion] (f2),
                };
            \end{feynman}
        \end{tikzpicture}
    \end{center}
    \caption{Higgsinos annihilate mainly through the $W$ portal as the coupling is unsuppressed. Here SM denotes Standard Model fermions.
    \label{fig:Higgsinoann}}
\end{figure}
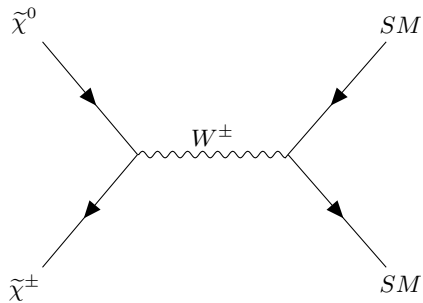

The Higgsino multiplet consists of two neutral Higgsinos and one charged Higgsino, all with masses close to $|\mu|$. All of them mix with gauginos, although the mixing is small if $|M_{1}|,|M_{2}|\gg |\mu|$. The annihilation of Higgsinos is mainly a coannihilation process of the neutral and charged Higgsino through the $W$ boson, shown in \cref{fig:Higgsinoann}. Since the coupling is the weak gauge coupling, the annihilation is always efficient.

A subleading contribution in the annihilation comes from the $Z$ portal. The coupling between the lightest Higgsinos and the $Z$ boson is
\begin{equation}\label{eq:Higgsinodd}
\lambda_{\tilde{\chi}\tilde{\chi}Z}=\frac{-ig\gamma^{5}}{2\cos\theta_{W}}(|N_{11}|^{2}-|N_{12}|^{2}).
\end{equation}
Here $N_{11}$ and $N_{12}$ denote the up- and down-type Higgsino components of the lightest neutralino. When $|M_{1}|,|M_{2}|\gg |\mu|$, the mass matrix of \cref{eq:Higgsinomass} leads to $|N_{11}|$ and $|N_{12}|$ being nearly equal ($\sim 1/\sqrt{2}$) and hence, in the case of an almost pure Higgsino, the coupling of \cref{eq:Higgsinodd} is strongly suppressed. As a result, the Higgsino produces a much weaker signal in direct-detection than one would expect from its annihilation rate.

However, even before the latest results on the spin-independent WIMP-nucleon cross section \cite{LZ:2022ufs,PandaX:2024qfu,XENON:2025vwd}, a large part of the parameter space, even with the suppressed couplings, was already excluded. The latest LZ Collaboration bound \cite{LZ:2024zvo} on the direct-detection cross section now requires the gauginos to be around $15$~TeV in order for the Higgsino couplings to the $Z$ boson to be sufficiently suppressed. Such a spectrum cannot be considered natural even though it is technically allowed.\footnote{The heavier Higgsinos would become long-lived, so further studies would be needed to evaluate to what extent the LHC searches of displaced vertices or disappearing tracks would have excluded such a scenario already. For the unphysical model of a pure Higgsino, the bound based on disappearing tracks is 210 GeV \cite{ATLAS:2022rme}.} These constraints are obviously evaded if the sneutrino is the dark matter candidate.

\subsection{Freeze-in component}

Right-handed sneutrinos can be produced through their interactions with both the Higgs sector and the Higgsino sector. As discussed in \cref{sec:model}, these interactions are proportional to the neutrino Yukawa couplings $y^{\nu}$ or the trilinear scalar couplings $A_{\nu}$. In what follows, we assume the latter to vanish. 

\begin{figure}
    \begin{subfigure}{0.8\textwidth}
        \begin{tabular}{c}
            \begin{tikzpicture}
                \begin{feynman}
                    \vertex (a);
                    \vertex [below = 1.0cm of a] (b);
                    \vertex [above = 0.6cm of a] (c);
                    \vertex [below = 0.6cm of b] (d);
    
                    \vertex [left = 0.6cm of c] (i1) {\(\widetilde{\nu}_L\)};
                    \vertex [right = 0.6cm of c] (f1) {\(\widetilde{\nu}_R\)};
                    \vertex [left = 0.55cm of d] (i2) {\shortstack{SM,$\widetilde{B}${,} \\$\widetilde{W}^\pm,\widetilde{W}^0$}};
                    \vertex [right = 0.55cm of d] (f2) {\shortstack{SM,$\widetilde{B}${,} \\$\widetilde{W}^\pm,\widetilde{W}^0$}};
                
                    \diagram*{
                      (i1) -- [scalar] (a) -- [scalar] (f1),
                      (a)  -- [scalar, edge label=\(h\)] (b),
                      (i2) -- [fermion] (b) -- [fermion] (f2),
                    };
                \end{feynman}
            \end{tikzpicture}
        \end{tabular}
        \begin{tabular}{c}
            \begin{tikzpicture}
                \begin{feynman}
                    \vertex (a);
                    \vertex [right = 1.0cm of a] (b);
                    \vertex [left = 0.6cm of a] (c);
                    \vertex [right = 0.6cm of b] (d);
    
                    \vertex [above = 0.6cm of d] (f1) {\(\widetilde{\nu}_R\)};
                    \vertex [below = 0.6cm of d] (f2) {\(\widetilde{\nu}_L\)};
                    \vertex [above = 0.55cm of c] (i1) {\shortstack{SM,$\widetilde{B}${,} \\$\widetilde{W}^\pm,\widetilde{W}^0$}};
                    \vertex [below = 0.55cm of c] (i2) {\shortstack{SM,$\widetilde{B}${,} \\$\widetilde{W}^\pm,\widetilde{W}^0$}};
                
                    \diagram*{
                      (i1) -- [fermion] (a) -- [fermion] (i2),
                      (a)  -- [scalar, edge label=\(h\)] (b),
                      (f1) -- [scalar] (b) -- [scalar] (f2),
                    };
                \end{feynman}
            \end{tikzpicture}
        \end{tabular}
        \begin{tabular}{c}
            \begin{tikzpicture}
                \begin{feynman}
                    \vertex (a);
                    \vertex [below = 1.0cm of a] (b);
                    \vertex [above = 0.7cm of a] (c);
                    \vertex [below = 0.7cm of b] (d);
    
                    \vertex [left = 0.7cm of c] (i1) {\(\widetilde{\nu}_L\)};
                    \vertex [right = 0.7cm of c] (f1) {\(\widetilde{\nu}_R\)};
                    \vertex [left = 0.7cm of d] (i2) {\(\widetilde{t}{,} \widetilde{\tau}\)};
                    \vertex [right = 0.7cm of d] (f2) {\(\widetilde{t}{,} \widetilde{\tau}\)};
                
                    \diagram*{
                      (i1) -- [scalar] (a) -- [scalar] (f1),
                      (a)  -- [scalar, edge label=\(h\)] (b),
                      (i2) -- [scalar] (b) -- [scalar] (f2),
                    };
                \end{feynman}
            \end{tikzpicture}
        \end{tabular}
        \begin{tabular}{c}
            \begin{tikzpicture}
                \begin{feynman}
                    \vertex (a);
                    \vertex [right = 1.0cm of a] (b);
                    \vertex [left = 0.7cm of a] (c);
                    \vertex [right = 0.7cm of b] (d);
    
                    \vertex [above = 0.7cm of d] (f1) {\(\widetilde{\nu}_R\)};
                    \vertex [below = 0.7cm of d] (f2) {\(\widetilde{\nu}_L\)};
                    \vertex [above = 0.7cm of c] (i1) {\(\widetilde{t}{,} \widetilde{\tau}\)};
                    \vertex [below = 0.7cm of c] (i2) {\(\widetilde{t}{,} \widetilde{\tau}\)};
                
                    \diagram*{
                      (i1) -- [scalar] (a) -- [scalar] (i2),
                      (a)  -- [scalar, edge label=\(h\)] (b),
                      (f1) -- [scalar] (b) -- [scalar] (f2),
                    };
                \end{feynman}
            \end{tikzpicture}
        \end{tabular}
        \caption{}
        \label{fig:neutralhiggsmediated}
    \end{subfigure}

    \medskip
    \begin{subfigure}{0.38\textwidth}
        \begin{tabular}{c}
            \begin{tikzpicture}
                \begin{feynman}
                    \vertex (a);
                    \vertex [below = 1.0cm of a] (b);
                    \vertex [above = 0.7cm of a] (c);
                    \vertex [below = 0.7cm of b] (d);
    
                    \vertex [left = 0.7cm of c] (i1) {\(\widetilde{l}^\pm\)};
                    \vertex [right = 0.7cm of c] (f1) {\(\widetilde{\nu}_R\)};
                    \vertex [left = 0.6cm of d] (i2) {\shortstack{t{,} b{,} \\ $\tau${,} $\nu_L$}};
                    \vertex [right = 0.6cm of d] (f2) {\shortstack{b{,} t{,} \\$\nu_L${,}$\tau$}};
                
                    \diagram*{
                      (i1) -- [scalar] (a) -- [scalar] (f1),
                      (a)  -- [scalar, edge label=\(H^\pm\)] (b),
                      (i2) -- [fermion] (b) -- [fermion] (f2),
                    };
                \end{feynman}
            \end{tikzpicture}
        \end{tabular}
        \begin{tabular}{c}
            \begin{tikzpicture}
                \begin{feynman}
                    \vertex (a);
                    \vertex [right = 1.0cm of a] (b);
                    \vertex [left = 0.7cm of a] (c);
                    \vertex [right = 0.7cm of b] (d);
    
                    \vertex [above = 0.7cm of d] (f1) {\(\widetilde{\nu}_R\)};
                    \vertex [below = 0.7cm of d] (f2) {\(\widetilde{\nu}_L\)};
                    \vertex [above = 0.6cm of c] (i1) {\shortstack{t{,} b{,} \\ $\tau${,} $\nu_L$}};
                    \vertex [below = 0.6cm of c] (i2) {\shortstack{b{,} t{,} \\$\nu_L${,}$\tau$}};
                
                    \diagram*{
                      (i1) -- [fermion] (a) -- [fermion] (i2),
                      (a)  -- [scalar, edge label=\(H^\pm\)] (b),
                      (f1) -- [scalar] (b) -- [scalar] (f2),
                    };
                \end{feynman}
            \end{tikzpicture}
        \end{tabular}
        \caption{}
        \label{fig:chargedhiggsmediated}
    \end{subfigure}
    \hfill
    \begin{subfigure}{0.61\textwidth}
        \begin{tabular}{c}
            \begin{tikzpicture}
                \begin{feynman}
                    \vertex (a);
                    \vertex [below = 1.0cm of a] (b);
                    \vertex [above = 0.7cm of a] (c);
                    \vertex [below = 0.7cm of b] (d);
    
                    \vertex [left = 0.7cm of c] (i1) {\(\nu_L\)};
                    \vertex [right = 0.7cm of c] (f1) {\(\widetilde{\nu}_R\)};
                    \vertex [left = 0.6cm of d] (i2) {\shortstack{$\widetilde{t}${,} $\widetilde{b}${,} \\ $\widetilde{\tau}${,}  $W^\pm$}};
                    \vertex [right = 0.6cm of d] (f2) {\shortstack{t{,} b{,} \\$\tau${,} $\widetilde{H}^\pm$}};
                
                    \diagram*{
                      (i1) -- [fermion] (a) -- [scalar] (f1),
                      (a)  -- [fermion, edge label=\(\widetilde{H}_{1,2}\)] (b),
                      (i2) -- [scalar] (b) -- [fermion] (f2),
                    };
                \end{feynman}
            \end{tikzpicture}
        \end{tabular}
        \begin{tabular}{c}
            \begin{tikzpicture}
                \begin{feynman}
                    \vertex (a);
                    \vertex [below = 1.0cm of a] (b);
                    \vertex [above = 0.7cm of a] (c);
                    \vertex [below = 0.7cm of b] (d);
    
                    \vertex [left = 0.7cm of c] (i1) {\(\nu_L\)};
                    \vertex [right = 0.7cm of c] (f1) {\(\widetilde{\nu}_R\)};
                    \vertex [left = 0.6cm of d] (i2) {\shortstack{t{,} b{,} \\$\tau${,} $\widetilde{H}^\pm$}};
                    \vertex [right = 0.6cm of d] (f2) {\shortstack{$\widetilde{t}${,} $\widetilde{b}${,} \\ $\widetilde{\tau}${,}  $W^\pm$}};
                
                    \diagram*{
                      (i1) -- [fermion] (a) -- [scalar] (f1),
                      (a)  -- [fermion, edge label=\(\widetilde{H}_{1,2}\)] (b),
                      (i2) -- [fermion] (b) -- [scalar] (f2),
                    };
                \end{feynman}
            \end{tikzpicture}
        \end{tabular}
        \begin{tabular}{c}
            \begin{tikzpicture}
                \begin{feynman}
                    \vertex (a);
                    \vertex [right = 1.0cm of a] (b);
                    \vertex [left = 0.7cm of a] (c);
                    \vertex [right = 0.7cm of b] (d);
    
                    \vertex [above = 0.7cm of d] (f1) {\(\widetilde{\nu}_R\)};
                    \vertex [below = 0.7cm of d] (f2) {\(\nu_L\)};
                    \vertex [above = 0.6cm of c] (i1) {\shortstack{t{,} b{,} \\$\tau${,} $\widetilde{H}^\pm$}};
                    \vertex [below = 0.6cm of c] (i2) {\shortstack{$\widetilde{t}${,} $\widetilde{b}${,} \\ $\widetilde{\tau}${,}  $W^\pm$}};
                
                    \diagram*{
                      (i1) -- [fermion] (a) -- [scalar] (i2),
                      (a)  -- [fermion, edge label=\(\widetilde{H}_{1,2}\)] (b),
                      (f1) -- [scalar] (b) -- [fermion] (f2),
                    };
                \end{feynman}
            \end{tikzpicture}
        \end{tabular}
        \caption{}
        \label{fig:neutralHiggsinomediated}
    \end{subfigure}

    \medskip
    \begin{subfigure}{0.59\textwidth}
        \begin{tabular}{c}
            \begin{tikzpicture}
                \begin{feynman}
                    \vertex (a);
                    \vertex [below = 1.0cm of a] (b);
                    \vertex [above = 0.7cm of a] (c);
                    \vertex [below = 0.7cm of b] (d);
    
                    \vertex [left = 0.7cm of c] (i1) {\(l^\pm\)};
                    \vertex [right = 0.7cm of c] (f1) {\(\widetilde{\nu}_R\)};
                    \vertex [left = 0.6cm of d] (i2) {\shortstack{$\widetilde{t}${,} $\widetilde{b}${,} \\ $\widetilde{\tau}${,}  $\widetilde{\nu}_L$}};
                    \vertex [right = 0.6cm of d] (f2) {\shortstack{b{,} t{,} \\$\nu_L${,} $\tau$}};
                
                    \diagram*{
                      (i1) -- [fermion] (a) -- [scalar] (f1),
                      (a)  -- [fermion, edge label=\(\widetilde{H}^\pm\)] (b),
                      (i2) -- [scalar] (b) -- [fermion] (f2),
                    };
                \end{feynman}
            \end{tikzpicture}
        \end{tabular}
        \begin{tabular}{c}
            \begin{tikzpicture}
                \begin{feynman}
                    \vertex (a);
                    \vertex [below = 1.0cm of a] (b);
                    \vertex [above = 0.7cm of a] (c);
                    \vertex [below = 0.7cm of b] (d);
    
                    \vertex [left = 0.7cm of c] (i1) {\(l^\pm\)};
                    \vertex [right = 0.7cm of c] (f1) {\(\widetilde{\nu}_R\)};
                    \vertex [left = 0.6cm of d] (i2) {\shortstack{b{,} t{,} \\$\nu_L${,} $\tau$}};
                    \vertex [right = 0.6cm of d] (f2) {\shortstack{$\widetilde{t}${,} $\widetilde{b}${,} \\ $\widetilde{\tau}${,}  $\widetilde{\nu}_L$}};
                
                    \diagram*{
                      (i1) -- [fermion] (a) -- [scalar] (f1),
                      (a)  -- [fermion, edge label=\(\widetilde{H}^\pm\)] (b),
                      (i2) -- [fermion] (b) -- [scalar] (f2),
                    };
                \end{feynman}
            \end{tikzpicture}
        \end{tabular}
        \begin{tabular}{c}
            \begin{tikzpicture}
                \begin{feynman}
                    \vertex (a);
                    \vertex [right = 1.0cm of a] (b);
                    \vertex [left = 0.7cm of a] (c);
                    \vertex [right = 0.7cm of b] (d);
    
                    \vertex [above = 0.7cm of d] (f1) {\(\widetilde{\nu}_R\)};
                    \vertex [below = 0.7cm of d] (f2) {\(l^\pm\)};
                    \vertex [above = 0.6cm of c] (i1) {\shortstack{b{,} t{,} \\$\nu_L${,} $\tau$}};
                    \vertex [below = 0.6cm of c] (i2) {\shortstack{$\widetilde{t}${,} $\widetilde{b}${,} \\ $\widetilde{\tau}${,}  $\widetilde{\nu}_L$}};
                
                    \diagram*{
                      (i1) -- [fermion] (a) -- [scalar] (i2),
                      (a)  -- [fermion, edge label=\(\widetilde{H}^\pm\)] (b),
                      (f1) -- [scalar] (b) -- [fermion] (f2),
                    };
                \end{feynman}
            \end{tikzpicture}
        \end{tabular}
        \caption{}
        \label{fig:chargedHiggsinomediated}
    \end{subfigure}
    
    \medskip
    \begin{subfigure}{0.20\textwidth}
        \begin{tabular}{c}
            \begin{tikzpicture}
                \begin{feynman}
                    \vertex (a);
                    \vertex [below = 1.0cm of a] (b);
                    \vertex [above = 0.7cm of a] (c);
                    \vertex [below = 0.7cm of b] (d);
    
                    \vertex [left = 0.7cm of c] (i1) {\(\widetilde{H}_{1,2}\)};
                    \vertex [right = 0.7cm of c] (f1) {\(\widetilde{\nu}_R\)};
                    \vertex [left = 0.7cm of d] (i2) {\(\tau\)};
                    \vertex [right = 0.7cm of d] (f2) {\(W^-\)};
                
                    \diagram*{
                      (i1) -- [fermion] (a) -- [scalar] (f1),
                      (a)  -- [fermion, edge label=\(\nu_L\)] (b),
                      (i2) -- [fermion] (b) -- [scalar] (f2),
                    };
                \end{feynman}
            \end{tikzpicture}
        \end{tabular}
        \caption{}
        \label{fig:neutrinomediated}
    \end{subfigure}
    \hspace{5mm}
    \begin{subfigure}{0.65\textwidth}
        \begin{tabular}{c}
            \begin{tikzpicture}
                \begin{feynman}
                    \vertex (a);
                    \vertex [below = 1.0cm of a] (b);
                    \vertex [above = 0.7cm of a] (c);
                    \vertex [below = 0.7cm of b] (d);
    
                    \vertex [left = 0.7cm of c] (i1) {\(h\)};
                    \vertex [right = 0.7cm of c] (f1) {\(\widetilde{\nu}_R\)};
                    \vertex [left = 0.6cm of d] (i2) {\shortstack{$\widetilde{\chi}^\pm_{1,2}{,}$ \\ $\widetilde{\chi}_{1,2,3,4}$}};
                    \vertex [right = 0.6cm of d] (f2) {\shortstack{$\tau{,}$ \\$\nu$}};
                
                    \diagram*{
                      (i1) -- [scalar] (a) -- [scalar] (f1),
                      (a)  -- [scalar, edge label=\(\widetilde{\nu}_L\)] (b),
                      (i2) -- [fermion] (b) -- [fermion] (f2),
                    };
                \end{feynman}
            \end{tikzpicture}
        \end{tabular}
        \begin{tabular}{c}
            \begin{tikzpicture}
                \begin{feynman}
                    \vertex (a);
                    \vertex [below = 1.0cm of a] (b);
                    \vertex [above = 0.7cm of a] (c);
                    \vertex [below = 0.7cm of b] (d);
    
                    \vertex [left = 0.7cm of c] (i1) {\(h\)};
                    \vertex [right = 0.7cm of c] (f1) {\(\widetilde{\nu}_R\)};
                    \vertex [left = 0.6cm of d] (i2) {\shortstack{$\tau{,}$ \\$\nu$}};
                    \vertex [right = 0.6cm of d] (f2) {\shortstack{$\widetilde{\chi}^\pm_{1,2}{,}$ \\ $\widetilde{\chi}_{1,2,3,4}$}};
                
                    \diagram*{
                      (i1) -- [scalar] (a) -- [scalar] (f1),
                      (a)  -- [scalar, edge label=\(\widetilde{\nu}_L\)] (b),
                      (i2) -- [fermion] (b) -- [fermion] (f2),
                    };
                \end{feynman}
            \end{tikzpicture}
        \end{tabular}
        \begin{tabular}{c}
            \begin{tikzpicture}
                \begin{feynman}
                    \vertex (a);
                    \vertex [right = 1.0cm of a] (b);
                    \vertex [left = 0.7cm of a] (c);
                    \vertex [right = 0.7cm of b] (d);
    
                    \vertex [above = 0.7cm of d] (f1) {\(\widetilde{\nu}_R\)};
                    \vertex [below = 0.7cm of d] (f2) {\(h\)};
                    \vertex [above = 0.6cm of c] (i1) {\shortstack{$\tau{,}$ \\$\nu$}};
                    \vertex [below = 0.6cm of c] (i2) {\shortstack{$\widetilde{\chi}^\pm_{1,2}{,}$ \\ $\widetilde{\chi}_{1,2,3,4}$}};
                
                    \diagram*{
                      (i1) -- [fermion] (a) -- [fermion] (i2),
                      (a)  -- [scalar, edge label=\(\widetilde{\nu}_L\)] (b),
                      (f1) -- [scalar] (b) -- [scalar] (f2),
                    };
                \end{feynman}
            \end{tikzpicture}
        \end{tabular}
        \caption{}
        \label{fig:sneutrinomediated}
    \end{subfigure}

    \caption{Potential t- and s-channel processes contributing to the freeze-in production of the right-handed sneutrinos. The mediator can be (a) the SM Higgs boson, (b) the charged Higgs boson, (c) a neutral Higgsino, (d) the charged Higgsino, (e) a left-handed neutrino, or (f) a left-handed sneutrino. SM stands for Standard Model particles ($t$, $b$, $\tau$, $W$, $Z$).}
    \label{fig:relevant_diagram}
\end{figure}
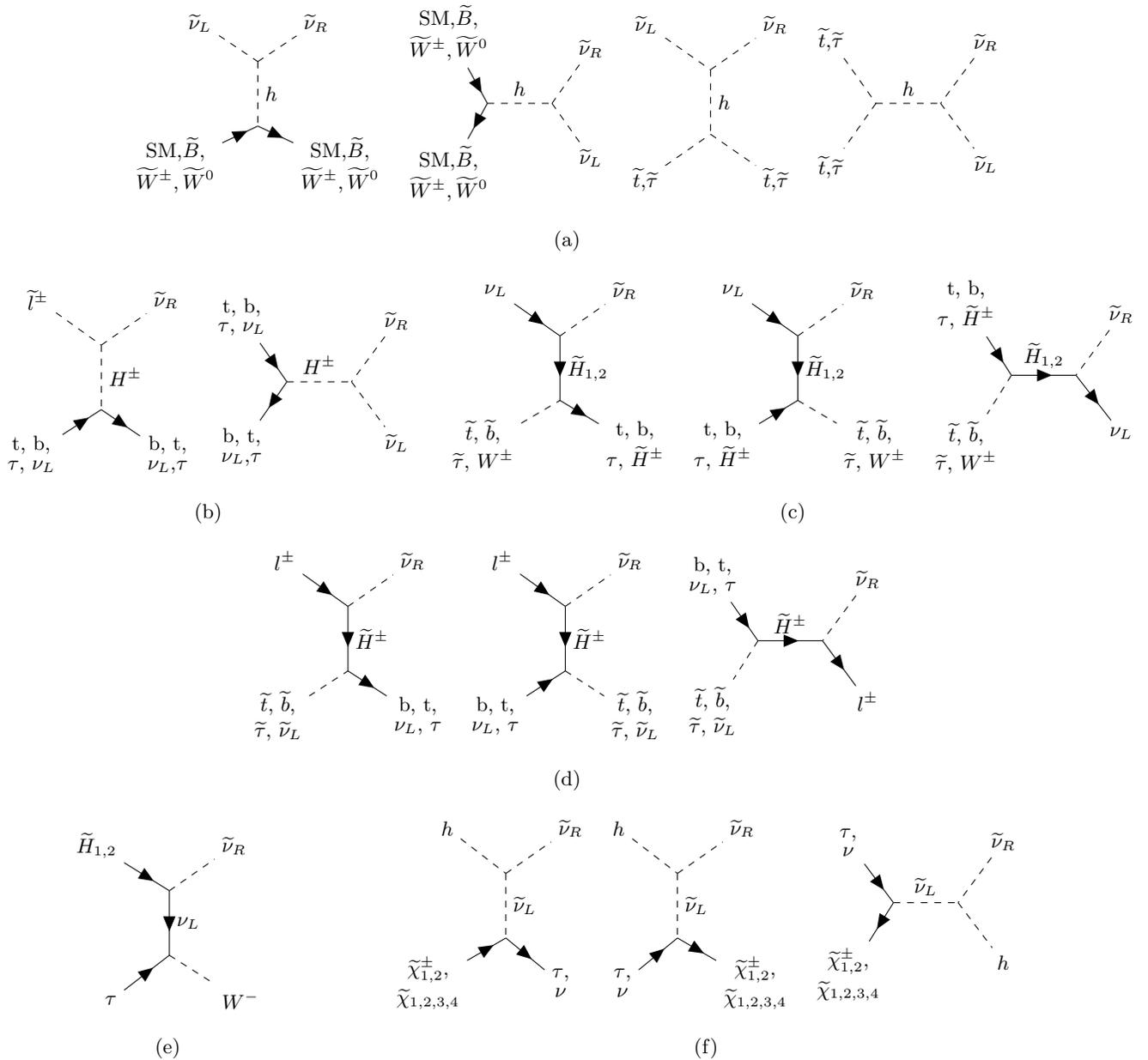

Since Higgs bosons and Higgsinos couple most strongly to the third generation of fermions, as well as to electroweak gauge bosons, and gauginos, the dominant freeze-in production arises from a limited set of $2\rightarrow 2$ processes as shown in \cref{fig:relevant_diagram}. Depending on the process, the mediator can appear in either the $t$ channel or the $s$ channel, although some of these processes are kinematically suppressed.

Broadly, two kinds of mediators are involved:
\begin{enumerate}
\item Higgs-sector mediators: These include both the neutral and charged Higgs bosons. Neutral Higgs exchange connects left-handed and right-handed sneutrinos to third-generation fermions or gauge bosons as shown in \cref{fig:neutralhiggsmediated}. Charged Higgs exchange instead involves charged sleptons and right-handed sneutrinos, again coupled to third-generation fermions in \cref{fig:chargedhiggsmediated}.
\item Higgsino-sector mediators: Neutral and charged Higgsinos can also mediate freeze-in production. Neutral Higgsino exchange couples left-handed neutrinos and right-handed sneutrinos to third-generation quark-squark or tau-stau pair (in \cref{fig:neutralHiggsinomediated}), while charged Higgsino exchange connects charged leptons and right-handed sneutrinos to third-generation quark-squark, tau-sneutrino or stau-neutrino pairs (in \cref{fig:chargedHiggsinomediated}).
\end{enumerate}

Left-handed neutrinos (\cref{fig:neutrinomediated}) and sneutrinos (\cref{fig:sneutrinomediated}) may also appear as mediators in some subdominant channels. In all cases, the resulting production rates are controlled by the smallness of the Yukawa couplings $y^\nu$, consistent with the freeze-in mechanism. 

\section{Results}\label{sec:results}
\subsection{Numerical procedure}

\renewcommand{\arraystretch}{1.2} 
\begin{table}
    \begin{tabular}{@{\hspace{8pt}}c@{\hspace{8pt}}|@{\hspace{8pt}}c@{\hspace{8pt}}|@{\hspace{8pt}}c@{\hspace{8pt}}|@{\hspace{8pt}}c@{\hspace{8pt}}}
        \hline
        \textbf{Parameter} & \textbf{BP1} & \textbf{BP2} & \textbf{Range} \\
        \hline
        $m_{\nu,11}$ [GeV] & 4.00E-09 & 3.00E-08 & [ 2.00E-09, 7.000E-09] \\
        $m_{\nu,22}$ [GeV] & 4.50E-09 & 2.50E-08 & [ 2.50E-09, 7.000E-09] \\
        $m_{\nu,33}$ [GeV] & 4.80E-09 & 4.00E-08 & [ 2.50E-09, 7.000E-09] \\
        $y^{\nu}_{11}$ & \,\,9.20E-13 & \,\,8.30E-13 & [ 7.40E-13, 1.130E-12] \\
        $y^{\nu}_{12}$ & \,\,9.30E-13 & \,\,8.50E-13 & [ 7.50E-13, 1.140E-12] \\
        $y^{\nu}_{13}$ &    -9.10E-13 &    -8.20E-13 & [-1.13E-12,-7.300E-13] \\
        $y^{\nu}_{21}$ & \,\,1.10E-12 & \,\,9.10E-13 & [ 9.10E-13, 1.330E-12] \\
        $y^{\nu}_{22}$ &    -9.80E-13 &    -8.90E-13 & [-1.24E-12,-7.600E-13] \\
        $y^{\nu}_{23}$ & \,\,1.30E-12 & \,\,9.30E-13 & [ 1.14E-13, 1.500E-12] \\
        $y^{\nu}_{31}$ & \,\,1.10E-12 & \,\,9.00E-13 & [ 9.00E-13, 1.340E-12] \\
        $y^{\nu}_{32}$ &    -9.90E-13 &    -8.80E-13 & [-1.25E-12,-7.600E-13] \\
        $y^{\nu}_{33}$ & \,\,1.30E-12 & \,\,9.30E-13 & [ 1.13E-13, 1.500E-12] \\
        $m_{\tilde{\chi}^0}$ [GeV] & 332.309 & 332.297 & - \\
        $m_{\tilde{\nu_{R1}}}$ [GeV] & 137.840 & 200.000 & - \\
        $m_{\tilde{\nu_{R2}}}$ [GeV] & 600.000 & 600.000 & - \\
        $m_{\tilde{\nu_{R3}}}$ [GeV] & 707.107 & 707.107 & - \\
        $m_{\nu_{R1}}$ [eV] & 4.018 & 2.822 & - \\
        $m_{\nu_{R2}}$ [eV] & 4.523 & 3.321 & - \\
        $m_{\nu_{R3}}$ [eV] & 4.826 & 3.621 & - \\
        \hline
    \end{tabular} 
    \caption{Parameter ranges used in our scans and two sample benchmark points to show our case. These benchmarks have sneutrino as LSP. Masses of the relevant particles are also mentioned at the end. In addition, we fix $\tan{\beta}=20$ and set $\mu,\,M_1,\,M_2, \text{and}~M_3~\text{to}~330, 880, 900~\text{and}~3500~\mathrm{GeV},$ respectively.}
    \label{tab:parameters}
\end{table}

\renewcommand{\arraystretch}{1} 

The model files were generated with \textsc{Sarah v4.14.5} \cite{Staub:2008uz,Staub:2013tta} and the particle spectrum was generated with \textsc{SPheno v4.0.5} \cite{Porod:2003um,Porod:2011nf}. The relic abundance calculations were performed using \textsc{micrOMEGAs v5.3.41}\cite{Belanger:2018ccd}. All our calculations were carried out in Feynman gauge. It is important to mention that \textsc{micrOMEGAs v 5.3.41} has errors in computing the correct freeze-in values in unitary gauge (via \texttt{ForceUG} option) and gives inconsistent results with increasing reheating temperature (\texttt{TR} option). We present two benchmark points, BP1 and BP2, where the sneutrino (our freeze-in candidate) appears as the LSP with different masses. The relevant parameters are listed in \cref{tab:parameters}.

Using BP1, we generated 800 points by randomly sampling the parameter range shown in \cref{tab:parameters}. In total, 12 parameters were varied, including $M^2_{\nu,N}$ and $y^\nu_{i,j}$. These ranges were chosen such that the total relic density remained within 10\% of the observed DM relic density. In addition, we ensured that the sum of the left-handed neutrino masses lie between $0.06~\mathrm{eV} \text{ and } 0.09~\mathrm{eV}$ \cite{DiValentino:2021hoh} for all benchmark points. \cref{fig:result} was generated using this dataset. Similarly, \cref{fig:mixingangle} was obtained from a dataset generated from a random scan of about 46,000 points without performing relic abundance calculations using \textsc{micrOMEGAs v 5.3.41}. We will discuss these figures in detail in the next section.

\subsection{Relic density}

\begin{figure}
    \centering
    \includegraphics[width=0.90\textwidth]{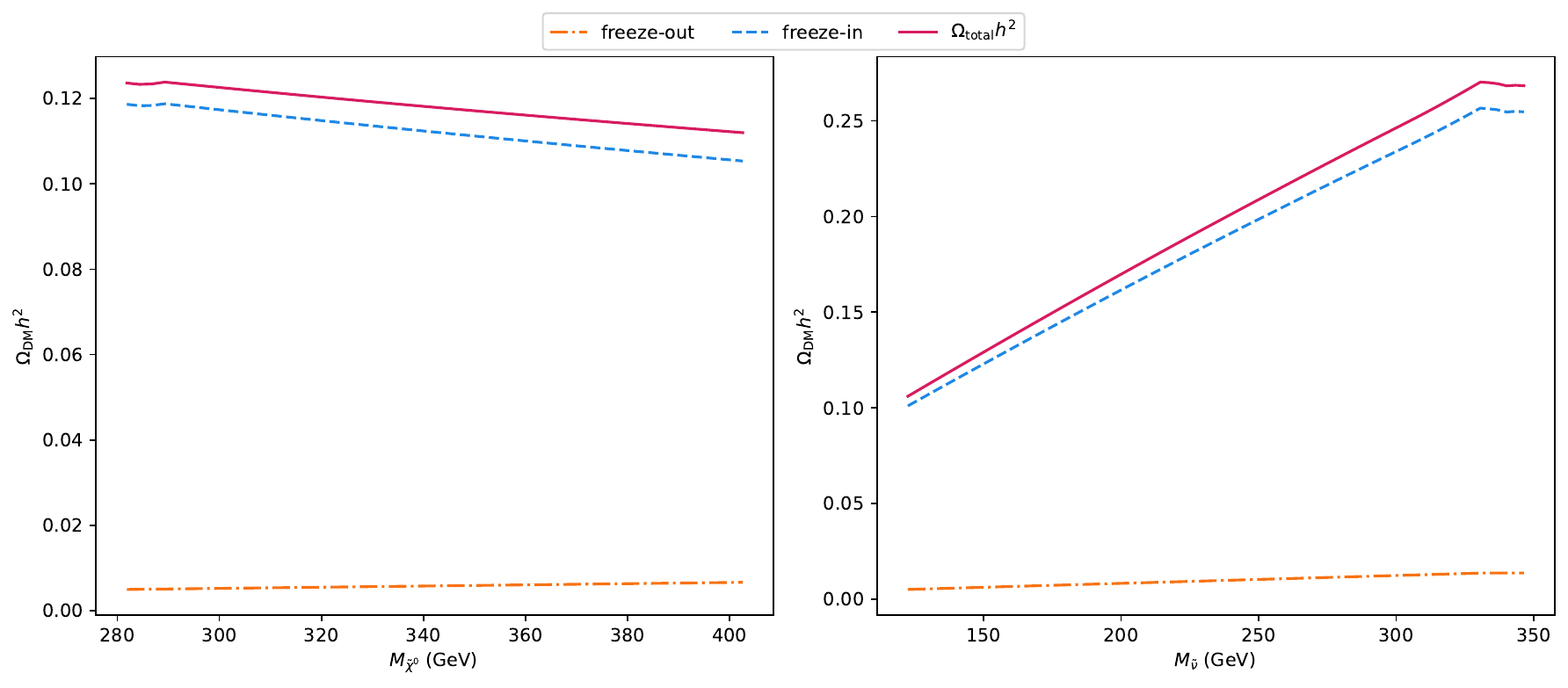}
    \caption{Relic density as a function of the scanned parameters $\mu$ and $M^2_{\nu,N}$ for BP1. The kinks indicate changes in the nature of the LSP from neutralino to sneutrino as we increase the parameter.}
    \label{fig:relicdensity}
\end{figure}

\begin{figure}
    \centering
    \includegraphics[width=0.90\textwidth]{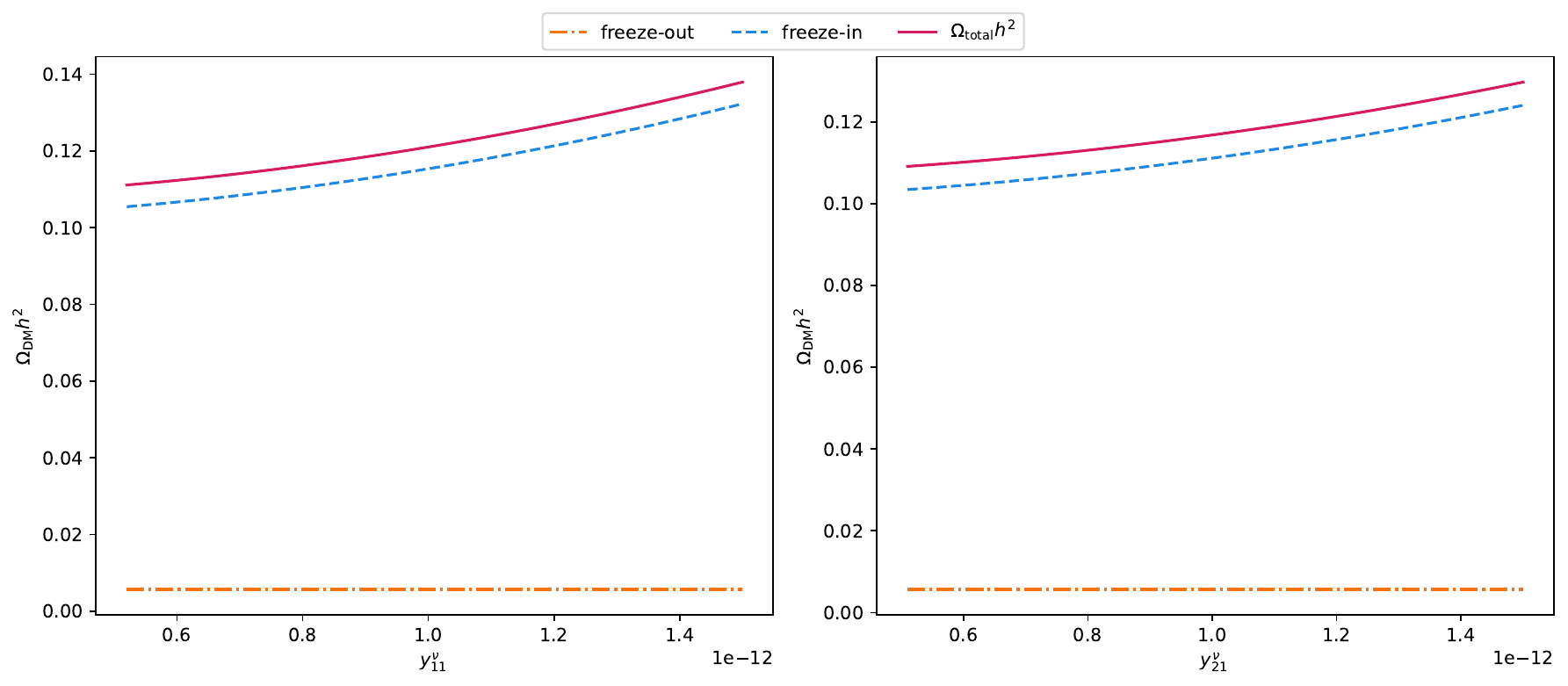}
    \caption{Relic density as a function of the scanned parameters $y^\nu_{1,1}$ and $y^\nu_{2,1}$, demonstrating a quadratic dependence across BP1.}
    \label{fig:relicdensityyukawa}
\end{figure}

Due to limitations in the total number of processes handled by the freeze-in module of \textsc{micrOMEGAs}, we manually added all relevant processes, as illustrated in \cref{fig:relevant_diagram}. For each generation of right-handed sneutrinos, there are about 7400 processes that in principle can produce them. However, not all of these are significant. After filtering, we retained a reduced list of about 700 processes per generation, which dominate the freeze-in production.

In the final scan, many processes involving neutrinos were excluded, as they required excessive computational time while contributing only marginally. In addition, we neglected any process with a contribution to the relic density below $10^{-6}$. Taken together, these contribute at most about $3 \times 10^{-3}$ to the relic density, corresponding to a maximum effect of about 3\%. For reference, we have taken the present DM relic abundance to be $0.1206 \pm 0.0021$ \cite{Planck:2018vyg}.

To study the parameter dependence in detail, we also performed a linear scan for BP1 as shown in \cref{fig:relicdensity,fig:relicdensityyukawa} by varying the Yukawa couplings and mass matrix entries within the same ranges listed in \cref{tab:parameters}. The resulting impact on the relic density is summarised above.

In \cref{fig:relicdensity}, the observed kinks indicate a change in the identity of the LSP. For the left panel, the region to the left of the kink corresponds to a neutralino LSP, whereas moving to the right of the kink shifts the LSP role to one of the sneutrinos. Similarly, in the right panel, the region to the left corresponds to a sneutrino LSP; as their masses increase, the LSP role transitions to the lightest Higgsino. In each case, the relic density shows a steady increase until the LSP identity changes, after which the contribution becomes nearly constant.

In \cref{fig:relicdensityyukawa}, the dependence of relic density on the Yukawa couplings exhibits a clearly parabolic behaviour, which is consistently visible across both panels. In our benchmarks, the freeze-out component accounts for about 5\% of the total relic abundance, while the remainder comes via freeze-in. This is mainly a consequence of the comparatively low Higgsino masses in our setup.

\subsection{Direct detection}

We also examined the possibility of a Higgsino-like dark matter candidate. However, we were unable to identify any viable benchmark point consistent with the relic density and existing constraints. Moreover, with the most recent bounds from direct-detection experiments \cite{LZ:2024zvo}, such scenarios are now entirely ruled out in the models discussed here. For the case where the sneutrino is an LSP, the spin-independent nucleon cross section is of the order $10^{-90} \text{cm}^2$, unobservable in direct-detection.

As discussed in \cref{sec:dmtheory}, Higgsino-like dark matter is excluded by direct-detection unless the gauginos are very heavy. In that case, the lightest neutralino is an almost pure Higgsino of the form $\tilde{\chi}^{0}_{1}\simeq \frac{1}{\sqrt{2}}(\tilde{H}^{0}_{u}-\tilde{H}^{0}_{d})$, for which the cancellation in \cref{eq:Higgsinodd} suppresses the direct-detection cross section. With the most recent bounds by the LZ experiment, the electroweak gauginos would need to have masses of at least $15$~TeV. Cases with very heavy gauginos have also been studied (see, \textit{e.g.}, \cite{Graham:2024syw}), as they provide a benchmark for inelastic direct-detection. However, such a heavy spectrum would lead to sizable loop-induced corrections to \cref{eq:MSSMEWSB} if we assume the gluino to lie in the same mass range as the electroweakinos (see, \textit{e.g.}, discussion in \cite{Baer:2015rja}). In addition, the loop corrections to the Higgs mass would tend to become too large.

\section{Sterile neutrino constraints}

\begin{figure}
    \centering
    \includegraphics[width=0.90\textwidth]{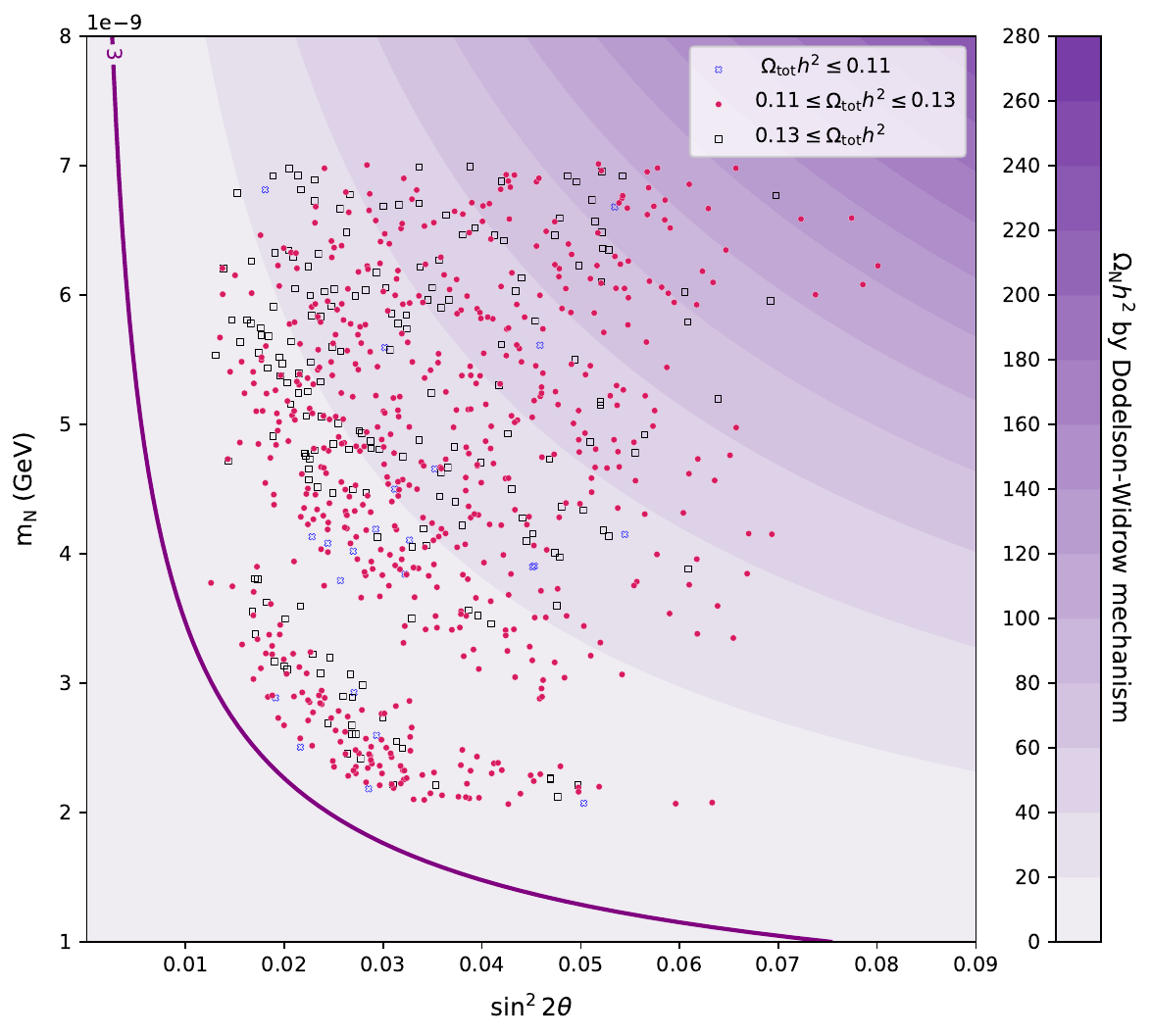}
    \caption{Mixing-angle–mass plane for a right-handed (sterile) neutrino: \(\sin^{2}\!\bigl(2\theta\bigr)\) vs \( m_{N}\,(\mathrm{GeV}) \). The model points are distributed based on the total relic abundance by freeze-in and freeze-out. The background colour map shows the sterile-neutrino relic abundance \( \Omega_{N}h^{2} \) from the Dodelson–Widrow mechanism.}
    \label{fig:result}
\end{figure}

\begin{figure}
    \centering
    \includegraphics[width=0.90\textwidth]{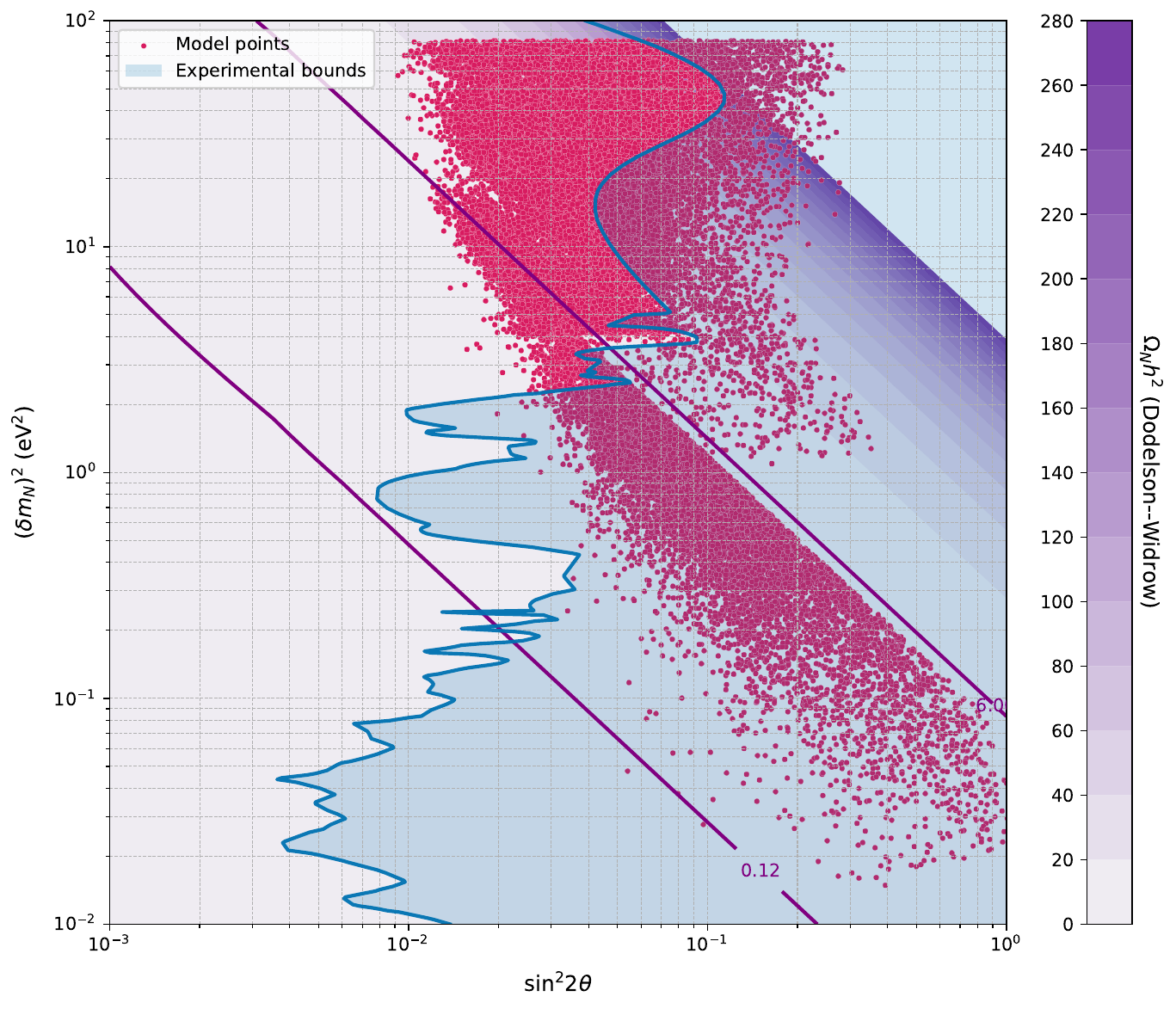}
    \caption{Exclusion in the \(\sin^{2}2\theta\)–\( (\delta m_{N})^{2} \) plane for a 3+1 oscillation framework. The blue contour shows the region excluded by experiments. Red points are random-scan model points that satisfy the measured sum of left-handed neutrino masses. The background shows the sterile–neutrino relic abundance \( \Omega_{N}h^{2} \) predicted via the Dodelson-Widrow mechanism, with reference isocontours at \(\Omega_{N}h^{2}=0.12\) and \( 6.00 \).}
    \label{fig:mixingangle}
\end{figure}

There are several direct and indirect constraints on right-handed (sterile) neutrinos. Collider searches \cite{CMS:2018iaf,ATLAS:2019kpx} and studies of kinematical end points in nuclear decays typically require large mixing between active and sterile neutrinos. Such a mixing cannot be realised within the type-I seesaw model and would require so large Yukawa couplings in the inverse and linear seesaw models that the sneutrino component would lead to overabundance in our scenario. In addition, short baseline oscillation experiments rule out very light sterile neutrinos below a few eV \cite{RENO:2020hva}. 

The indirect constraints are more severe. Light sterile neutrinos are relatively long-lived. They decay to active neutrinos via $N\rightarrow 3\nu$ if no other channels are kinematically available. The consequences of this decay depend on the sterile-neutrino lifetime. If the lifetime is of the order of one second, the decays will produce active neutrinos carrying energies higher than those expected from the thermal distribution, which can have an impact on Big Bang nucleosynthesis (BBN) \cite{Ruchayskiy:2012si}. Successful BBN therefore requires the sterile neutrinos either to decay well before BBN, allowing neutrinos to thermalise, or to be sufficiently long-lived.

If the sterile neutrinos have a lifetime longer than the age of the Universe, they constitute a (decaying) part of the dark matter. The dominant mechanism for sterile-neutrino dark matter generation is oscillations between active and sterile neutrinos known as the Dodelson--Widrow mechanism \cite{Dodelson:1993je}. In this case, the relic density depends primarily on the sterile-neutrino mass and the mixing angle between the active and sterile flavours. The relic density can be approximately given by \cite{Abazajian:2005gj,Dasgupta:2021ies}
\begin{equation}\label{eq:sterilerelic}
\Omega h^{2}\simeq 0.12 \left( \frac{m_{N}}{1\,\mathrm{keV}} \right)^{2}\left(\frac{\sin^{2}2\theta}{7.3\times 10^{-8}} \right)^{1.23},
\end{equation}
where $m_{N}$ is the sterile-neutrino mass and $\theta$ the mixing angle between the active and sterile neutrinos. The relic density depends rather mildly on model parameters since increasing the sterile-neutrino mass leads to a smaller mixing angle.

For all our benchmarks, we find that the lifetime of sterile neutrinos is of the order $10^{42}$\,GeV$^{-1}$ ($2.087\times 10^{10}$\,yr), which is more than the age of the Universe ($1.38\times10^{10}$\,yr). Thus, they are stable on cosmological scales and contribute to the DM abundance. 

As discussed earlier, the production of the right amount of sneutrino dark matter through freeze-in requires the neutrino Yukawa couplings of the order $\mathcal{O}(10^{-12})$. \cref{eq:seesaw} implies that $m_{N}=(y^{\nu}v\sin\beta)^2/2m_{\nu}$. The heaviest active neutrino must be at least $0.04$~eV which leads to sterile-neutrino masses in the range $5$--$20$~eV, not much above the lowest allowed values by short baseline oscillation experiments. The associated Dirac mass term is of the order of one electronvolt which leads to an active-sterile mixing angle $\theta\gtrsim 10^{-2}$.

Linear and inverse seesaw models allow us to suppress this mixing by letting the sterile neutrinos be more massive. In the linear seesaw model, however, this comes with a price as this introduces additional interactions between the sterile sneutrinos and SM particles, which makes the freeze-in contribution too large as discussed earlier.

In the inverse seesaw model, the mass matrix has the form
\begin{equation}
    m_{\mathrm{IS}}=
    \begin{pmatrix}
        0 & y^{\nu}v_{u} & 0\\
        y^{\nu}v_{u} & 0 & M_{N}\\
        0 & M_{N} & \mu_{S}
    \end{pmatrix}.
\end{equation}
If we wish to have heavier sterile neutrinos while keeping the light neutrino mass constant, the factor $\mu_{S}/M_{N}$ must be greater than 1. The more we increase the sterile-neutrino mass, the larger $\mu_{S}$ must become. For proper suppression of the mixing $\mu_{S}/M_{N}\gtrsim 5$ would be required. When $\mu_{S}$ is significantly larger than $M_{N}$, the lower $2\times 2$ block of the inverse seesaw mass matrix becomes equivalent to that of the type-I seesaw model. Then we get sterile-neutrino masses $M_{N}^{2}/\mu_{S}$ for $N$ and $\mu_{S}$ for $S$, which means that we end up having the top $2\times 2$ block and type-I seesaw model. Hence, the inverse seesaw model too cannot successfully suppress the Dodelson--Widrow contribution to the relic density. The only model where there is no problem is that of Dirac neutrinos, since there are no right-handed neutrinos.

In \cref{fig:result}, we plot the mass of the sterile neutrinos (with the largest mixing angle) against $\sin^{2}2\theta$ using points from the type-I seesaw model. The red dots correspond to points that yield the correct relic density with an error margin of $10\%$ accounting for numerical or package-related uncertainties. The black square and blue cross indicate benchmarks that lead to DM overabundance and underabundance, respectively. As the figure shows, all the points predict a relic density greater than 3, $\Omega_{N}h^{2}>3$, denoted by the reference line.

Another feature visible in \cref{fig:result} is the parabolic void in the lower left region of the plot, due to the lower bound on the Yukawa coupling specified in \cref{tab:parameters}. By reducing this lower limit by half, this parabolic void disappears. Nonetheless, such points are excluded since they predict excessive contribution to DM relic density from sterile neutrinos via the Dodelson–Widrow mechanism.

The overabundance is mainly driven by the heaviest right-handed neutrino. For example, $m_{N}=50$~eV and $\sin^{2}2\theta\gtrsim 10^{-4}$ yields $\Omega h^{2}\gtrsim 2$. In the type-I seesaw, it is impossible to suppress this mixing while preserving a viable neutrino mass spectrum. \cref{fig:mixingangle} illustrates this through a numerical scan, where the Yukawa coupling values are now relaxed to $10^{-15}$ and the sterile-neutrino masses down to $0.1~\mathrm{eV}$. Points that allow smaller dark matter contributions from the right-handed neutrinos are excluded by oscillation experiments (lower right region), while the allowed points lead to a clear overabundance (upper middle). The experiments included in the blue contour are Bugey-3\cite{Declais:1994su}, NEOS\cite{NEOS:2016wee}, DANSS\cite{Shitov:2019hli}, Daya Bay\cite{DayaBay:2024nip}, PROSPECT-I\cite{PROSPECT:2024gps} and KATRIN\cite{KATRIN:2025lph}.

In summary, although the combination of freeze-out and freeze-in can yield an acceptable relic density, the sterile-neutrino contribution is always too large. If sterile neutrinos were made sufficiently heavy with a sufficiently short lifetime, the couplings between the SM and sneutrinos would become too large to be viable.

\section{Conclusions}

We studied electroweak-scale seesaw models in a supersymmetric setting and examined the possibility of saturating the dark matter relic density through a combination of a traditional freeze-out dark matter candidate and the freeze-in of feebly coupled sneutrinos.

We showed that an acceptable relic density requires very small neutrino Yukawa couplings and a sterile-neutrino mass scale so low that the sterile neutrinos become cosmologically stable. In the region of parameter space where the relic density from the superpartners is viable, the production of sterile neutrinos in the early Universe via the Dodelson--Widrow mechanism leads to overabundance. Therefore, a scenario with freeze-in dark matter from sterile sneutrinos at the electroweak scale does not work for type-I, inverse, or linear seesaw mechanisms. For Dirac neutrinos, however, this scenario remains possible, as there are no sterile-neutrino mass eigenstates that would lead to overabundance.

\section*{Acknowledgments}
T.G. acknowledges Magnus Ehrnrooth Foundation for funding his doctoral studies. M.H. acknowledges the support from the Research Council of Finland (Grants \#342777 and No.\#371542). H.W. acknowledges the support from Carl Trygger Foundation (Grant CTS18:164) and Ruth and Nils-Erik Stenb\"ack Foundation. H.W. also thanks Helsinki Institute of Physics for hospitality during this project. The authors wish to thank the Finnish Computing Competence Infrastructure (FCCI) for supporting this project with computational and data storage resources.

\bibliography{bibliography} 

\end{document}